\def\Box{\kern1pt\vbox{\hrule height 1.2pt\hbox{\vrule width 1.2pt\hskip 3pt
   \vbox{\vskip 6pt}\hskip 3pt\vrule width 0.6pt}\hrule height 0.6pt}\kern1pt}

\documentclass[12pt]{article}
\usepackage{amssymb,amsmath}

\begin{document}
\begin{titlepage}
\begin{flushright}
UFIFT-HEP-03-19
\end{flushright}
\vspace{.4cm}
\begin{center}
\textbf{The Force of Gravity from a Lagrangian containing Inverse
  Powers of the Ricci Scalar}
\end{center}
\begin{center}
M. E. Soussa$^{\dagger}$ and R. P. Woodard$^{\ddagger}$\\
\vspace{.2cm}
\textit{Department of Physics \\ University of Florida \\
Gainesville, FL 32611 USA}
\end{center}
\begin{center} 
ABSTRACT
\end{center}

We determine the gravitational response to a diffuse source, in a
locally de Sitter background, of a class of theories which modify the 
Einstein-Hilbert action by adding a term proportional to an inverse 
power of the Ricci scalar. We find a linearly growing force which is
not phenomenologically acceptable.
\begin{flushleft}
\end{flushleft}
\vspace{.4cm}
\begin{flushleft}
$^{\dagger}$ e-mail: soussam@phys.ufl.edu \\
$^{\ddagger}$ e-mail: woodard@phys.ufl.edu
\end{flushleft}
\end{titlepage}

\section{Introduction}
Observations of Type Ia Supernovae give compelling evidence that
the universe is entering a phase of acceleration \cite{SN}. This means 
the current energy profile is dominated by an unknown source contributing 
negative pressure which has become known as ``dark energy''. When various 
recent data sets are combined, the fraction of dark energy present in the
universe is determined to have the value, $\Omega_\Lambda \simeq 0.73$
\cite{WMAP}. The number and quality of the most recent measurements
\cite{SN} leave little doubt that we are facing a real effect which
must be explained.

There has been no lack of theoretical effort to account for and describe
dark energy origins and dynamics \cite{DE1}-\cite{DE7}. 
None of the suggestions is without problems. A bare cosmological
constant will work, but one has to understand both why it is more than
120 orders of magnitude smaller than
its seemingly natural scale, and why it has just achieved dominance in the
current epoch \cite{SW,SC}. Quintessence based on a scalar field will
also work \cite{Q1}-\cite{Q5} but one must understand why it is homogeneous
\cite{VT} and again why it has achieved dominance now. Long range forces
\cite{LF1} and even quantum effects \cite{QE} have also been suggested.

A recent paper by Carroll, Duvvuri, Trodden, and Turner proposed a 
purely gravitational approach \cite{Carroll}. Late time acceleration is 
achieved by considering a subset of nonlinear gravity theories in which a
function of the Ricci scalar is added to the usual Einstein-Hilbert action,
\begin{equation}
S[g] = \frac1{16 \pi G} \int d^4x\sqrt{-g} \, [R + f(R)] . \label{mod}
\end{equation}
In the case of \cite{Carroll} this function was an inverse power,
\begin{equation}
f(R) = -\mu^{2(p+1)}R^{-p},\label{inversetheories}
\end{equation}
where $p>0$ and $\mu$ is an \textit{a priori} unknown parameter.
It has been argued that such inverse powers of $R$ may be closely
related with braneworlds and string theory \cite{Nojiri1}. By 
considering the standard Friedmann-Robertson-Walker cosmology,
Carroll, Duvvuri, Trodden and Turner showed that if $\mu \sim 
10^{-33}$eV (the inverse age of the universe) the equation of state 
parameter could fall within the range $-1<w_{\rm eff}\leq -2/3$ at 
late times.

Although higher derivatives typically bring negative energy degrees of
freedom, endowing the Lagrangian with nonlinear functions of the Ricci
scalar can sometimes be acceptable \cite{Strominger}. This will only
give rise to a single, spin zero higher derivative degree of freedom.
Since the lower derivative spin zero field is a constrained, negative
energy degree of freedom (the Newtonian potential), its higher derivative
counterpart can sometimes carry positive energy.  

Recent papers have considered different aspects of the model. Dick
considered the Newtonian limit in perturbation theory about a maximally 
symmetric background \cite{Dick}. Dolgov and Kawasaki discovered an 
apparently fatal instability in the interior of a matter distribution 
\cite{Dolgov}. However, Nojiri and Odintsov have shown that an $R^2$
can be added to the Lagrangian without changing the cosmological
solution, and that the coefficient of this term can be chosen to enormously
increase the time constant of the interior instability \cite{Nojiri2}.
Meng and Wang have explored perturbative corrections to cosmology \cite{MW}. 
Others have drawn connections with a special class of scalar-tensor 
theories \cite{Chiba}-\cite{Cline}.

Our own work concerns the response to a diffuse spherical matter
source after the epoch of acceleration has set in. The procedure will
be to solve for the perturbed Ricci scalar, whence we determine the 
gravitational force carried by the trace of the metric perturbation. We 
constrain the matter distribution to have the property that its rate of 
gravitational collapse is identical to the rate of spacetime
expansion, thereby fixing the \textit{physical} radius of the
distribution to a constant value. Further, we impose the condition
that inside the matter distribution the density is low enough to
justifiably employ a locally de Sitter background, in which case the
Ricci scalar can be solved exactly and remains constant.     

Our de Sitter background was considered by Carroll {\it et al.}
\cite{Carroll} and was found to be unstable.  However, the decay time 
$\sim \mu^{-1}$ is far too long to create any practical concern.  
The next section describes the calculation and shows the solution
outside the matter distribution to grow linearly.  We conclude by
remarking on the implications of this result.

\section{The Gravitational Response}
\noindent
We shall consider a gravitational action parameterized by $p > 0$,
\begin{equation}
S[g]=\frac1{16 \pi G}
\int d^4x\sqrt{-g}\left[R - \mu^{2(p+1)}R^{-p}\right]. \label{action}
\end{equation}
(We employ a spacelike metric with Ricci tensor $R_{\mu\nu} \equiv
\Gamma^{\rho}_{~\nu\mu , \rho} - \Gamma^{\rho}_{~ \rho \mu , \nu} +
\Gamma^{\rho}_{~ \rho \sigma} \Gamma^{\sigma}_{~ \nu\mu} - 
\Gamma^{\rho}_{~ \nu\sigma} \Gamma^{\sigma}_{~\rho \mu}$.)
Functionally varying with respect to the metric and setting it equal
to the matter stress energy tensor leads to the equations of motion,
\begin{multline}
\left[1+p\mu^{2(p+1)}R^{-(p+1)}\right]R_{\mu\nu} 
-\frac12\left[1-\mu^{2(p+1)}R^{-(p+1)}\right]Rg_{\mu\nu} \\
+p\mu^{2(p+1)}(g_{\mu\nu}\Box-D_\mu D_\nu)R^{-(p+1)} =
8 \pi G T_{\mu\nu}.\label{eom}
\end{multline}
$D_\mu$ is the covariant derivative and 
$\Box\equiv(-g)^{-1/2}\partial_\mu(\sqrt{-g}g^{\mu\nu}\partial_\nu)$ is
the covariant d'Alembertian.

Although one must really solve all components of the field equations 
(\ref{eom}) we can get an important part of the gravitational response 
by simply taking the trace. We shall also restrict to $p = 1$ for 
simplicity. Inside the matter distribution the trace equation is,
\begin{equation}
-R+3\frac{\mu^4}{R}+3\mu^4\Box \frac{1}{R^2} = 8 \pi G g^{\mu\nu} T_{\mu\nu}
\equiv \mathcal{T}. \label{trace}
\end{equation}
(Note that $\mathcal{T}$ is negative.) Normally, one would expect the 
matter stress energy to be redshifted by powers of the scale factor in 
an expanding universe. However, recall that this matter distribution 
possesses a rate of gravitational collapse equal to the rate of universal 
expansion, and thus $\mathcal{T}$ remains \textit{constant}. Since our 
matter source is also diffuse, we may perturb around a locally de Sitter 
space. For the interior solution, we are able to solve for $R$ exactly using 
equation (\ref{trace}) for the case $\mathcal{T}$ is constant and $D_\mu R=0$,
\begin{equation}
R_{\rm in}=-\frac{\mathcal{T}}2 \left[1\mp\sqrt{1+\frac{12\mu^4}{
\mathcal{T}^2}}\right].
\end{equation}
Obtaining de Sitter background obviously selects the negative root.
Further, we concentrate on the situation $\vert \mathcal{T} \vert \ll\mu^2$,
\begin{equation}
R_{\rm in}=\sqrt{3}\mu^2-\frac{\mathcal{T}}2+\cdots.
\end{equation} 

Outside the matter source we perturb around the de Sitter vacuum
solution,
\begin{equation}
R_{\rm out}=\sqrt3 \mu^2+\delta R. \label{R_out}
\end{equation}
Substituting (\ref{R_out}) into (\ref{trace}) and expanding to first
order in $\delta R$ yields the equation defining the Ricci scalar correction, 
\begin{equation}
\Box \delta R(x) +\sqrt3\mu^2\delta R(x) =  0. \label{pe1}
\end{equation}
In our locally de Sitter background the invariant length element is,
\begin{align}
ds^2 & \equiv -dt^2+a^2(t)d\vec{x}\cdot d\vec{x} \label{metric},\\
\intertext{with $a(t)$ having the property,}
H &\equiv \frac{\dot a}{a}={\rm constant}.
\end{align}
We can relate the Hubble constant $H$ to the parameter $\mu$ via the
vacuum Ricci scalar,
\begin{equation}
R = 12H^2  = \sqrt3\mu^2.
\end{equation}
Identifying $\Box=a^{-3}\partial_\rho(a^3g^{\rho\sigma}\partial_\sigma)$, 
we expand (\ref{pe1}),
\begin{equation}
\left[\partial^2 - 3H\partial_0+12H^2\right] \delta R(t,\vec{x})
= 0, 
\label{pe2} 
\end{equation}
where $\partial^2 \equiv -\partial_0^2+a^{-2}\nabla^2$.  It is evident
from (\ref{pe2}) that the frequency term has the wrong sign for
stability \cite{Carroll}. However, since the decay time is proportional to
$1/H$, we may safely ignore this issue.  

Seeking a solution of the form $\delta R = \delta
R(H a\Vert\vec{x}\Vert)$ allows us to convert (\ref{pe2}) into an
ordinary differential equation,
\begin{equation}
\left[(1-y^2)\frac{d^2}{dy^2}+\frac2y(1-2y^2)\frac{d}{dy}+12\right]
\delta R = 0,
\end{equation}
where $y\equiv Ha\Vert\vec{x}\Vert$.
To solve this equation we try a series of the form,
\begin{equation}
f_{\alpha}(y) = \sum_{n=0}^\infty f_n y^{\alpha+n}.
\end{equation}
Substituting this series into (\ref{pe2}) yields a solution with 
$\alpha =0$,
\begin{equation}
f_0(y) \equiv \sum_{n=0}^\infty \frac{\Gamma(n+\frac34-\frac{\sqrt{57}}{4}) 
\Gamma(n+\frac34+\frac{\sqrt{57}}{4})}{\Gamma(\frac34-\frac{\sqrt{57}}{4})
\Gamma(\frac34+\frac{\sqrt{57}}{4})}\frac{(2y)^{2n}}{(2n+1)!} ,
\end{equation}
and a solution with $\alpha = -1$,
\begin{equation}
f_{-1}(y) \equiv \frac1{y} \sum_{n=0}^\infty \frac{\Gamma(n+\frac14-
\frac{\sqrt{57}}{4}) \Gamma(n+\frac14+\frac{\sqrt{57}}{4})}{\Gamma(\frac14-
\frac{\sqrt{57}}{4}) \Gamma(\frac14+\frac{\sqrt{57}}{4})}\frac{(2y)^{2n}}{
(2n)!}.
\end{equation}
Both solutions converge for $0 < y < 1$. Both also have a logarithmic 
singularity at $y = 1$, which corresponds to the Hubble radius. We can
therefore employ them quite reliably within the visible universe.

The solution we seek is a linear combination,
\begin{equation}
\delta R(y) = \beta_1 f_0(y) + \beta_2 f_{-1}(y), \label{ext}
\end{equation}
whose coefficients are determined by the requirements that $\delta R(y)$
and its first derivative are continuous at the boundary of the matter
distribution. We employ a spherically symmetric distribution of matter, 
centered on the co-moving origin. If the matter distribution collapses at 
the same rate as the expansion of the universe, its physical radius is a 
constant we call $\rho$. (This means that the co-moving coordinate radius
is $\rho/a(t)$.) If the total mass of the distribution is $M$ we can
identify $\mathcal{T}$ as the constant,
\begin{equation}
\mathcal{T} = - \frac{8 \pi G M}{\frac43 \pi \rho^3} = - \frac{6 G M}{\rho^3} .
\end{equation}
In terms of our variable $y = H a(t) \Vert \vec{x} \Vert$, the boundary
of the matter distribution is at $y_0 = H \rho$. Demanding continuity of
the Ricci scalar and its first derivative at $y_0$ gives the following
result for the combination coefficients of the exterior solution (\ref{ext}),
\begin{align}
\beta_1 &= \frac{3MG}{\rho^3}\left[f_0(y_0)-\frac{f_0'(y_0)}{f_{-1}'(y_0)}
f_{-1}(y_0)\right]^{-1},\\
\beta_2 &= \frac{3MG}{\rho^3}\left[f_{-1}(y_0)-\frac{f_{-1}'(y_0)}{f_0'(y_0)}
f_0(y_0)\right]^{-1},
\end{align}
where a prime represents the derivative with respect to the argument.

We are now in a position to calculate the gravitational force carried by the
trace of the graviton field. The metric perturbation modifies the invariant 
length element as follows,
\begin{equation}
ds^2=-(1-h_{00})dt^2 + 2 a(t) h_{0i} dt dx^i + a^2(t) (\delta_{ij}+h_{ij})
dx^i dx^j .
\end{equation}  
Further defining $h\equiv -h_{00}+h_{ii}$ and imposing the gauge
condition,
\begin{equation}
h_{\mu\nu}^{\ \ ,\nu}-\frac12h_{,\mu}+3h^\nu_{\ \mu}(\ln a)_{, \nu}
= 0,
\end{equation}
allows us to express the Ricci scalar in terms of $h$,
\begin{equation}
\delta R = \frac12\biggl(-\partial^2h+ 4H\partial_0h\biggr).
\end{equation}
(Recall that we define $\partial_{\mu} \equiv (\partial_t,a^{-1} 
\vec{\nabla})$.) Assuming $h=h(y)$ as we did for $\delta R$ gives the 
equation for the gravitational force carried by $h$,
\begin{equation}
\left[(y^2-1)\frac{d}{dy}+\frac1y(5y^2-2)\right]h'(y)= \frac{2\delta R(y)}{H^2}
. \label{force}
\end{equation}
The solution to (\ref{force}) is,
\begin{equation}
h'(y)=-\frac{2}{y^2(1-y^2)^{3/2}}\int_{0}^y dy' y^{'2} 
(1-y^{'2})^{1/2} \frac{\delta R(y')}{H^2} . \label{hsol}
\end{equation}

At this point it is useful to consider the $y$ values which are relevant.
The Hubble radius corresponds to $y = 1$, whereas the typical distance between
galaxies corresponds to about $y = 10^{-4}$, and a typical galaxy radius 
would be about $y = 10^{-6}$. We are therefore quite justified in assuming
that $y_0 \ll 1$, and in specializing to the case of $y_0 \ll y \ll 1$. Now
consider the series expansions,
\begin{align}
f_0(y) &= 1-2y^2+\frac15 y^4 + \mathcal{O}(y^6) \qquad \ \ ,    
\beta_1 = \frac{3MG}{\rho^3}+\mathcal{O}(y_0^2), \\
f_{-1}(y)&=\frac1y\left[1-7y^2+\frac{14}{3}y^4+\mathcal{O}(y^6)\right], 
\beta_2 = -\frac{12MGy_0^3}{\rho^3}+\mathcal{O}(y_0^5).
\end{align}
We see first that $\vert \beta_2 \vert \ll \beta_1$ --- which means 
$\delta R(y) \approx \beta_1 f_0(y)$ --- and second, that $f_0(y) \sim 1$ 
--- which implies $\delta R(y) \approx -\mathcal{T}/2$. This means that 
the integrand in (\ref{hsol}) fails to fall off for $y > y_0$, so the integral 
continues to grow outside the boundary of the matter distribution. For 
small $y \gg y_0$ we have,
\begin{equation}
h'(y) = -\frac{2 G M}{H^2 \rho^3}y+\mathcal{O}(y^3). \label{solution}
\end{equation}

To see that this linear growth is a phenomenological disaster it suffices
to compare (\ref{solution}) with the result that would follow for the same
matter distribution, in the same locally de Sitter background, if the theory
of gravity had been general relativity with a positive cosmological constant 
$\Lambda = 3 H^2$. In that case $\delta R(y) = -\mathcal{T} \theta(y_0 - y)$ 
and, for $y > y_0$, the integral in (\ref{hsol}) gives,
\begin{eqnarray}
h'(y) \Bigl\vert_{GR} & = & \frac{\mathcal{T}}{4 H^2 y^2 (1 - y^2)^{\frac32}} 
\Bigl\{ {\rm arcsin}(y_0) - y_0 (1 - 2 y_0^2) \sqrt{1 - y_0^2} \Bigr\} . \\
& = & - \frac{4 G M H}{y^2} + \mathcal{O}(1).
\end{eqnarray}
The linear force law (\ref{solution}) of modified gravity is stronger by a
factor of $\frac12 (\frac{y}{y_0})^3$. For the force between two galaxies
this factor would be about a million.

\section{Conclusion and Remarks}
We have determined the gravitational response to a diffuse matter
source in a locally de Sitter background.  Our result is the leading
order result in the expansion variable $y$, the fractional Hubble
distance.  Equation (\ref{solution}) clearly forces us to disregard
the class of theories considered here (\ref{inversetheories}) when
compared to GR with a cosmological constant (for example, the
correction to the gravitational force between the
Milky Way and Andromeda increases by six orders of magnitude).  

The two assumptions made in our analysis were: 
\begin{itemize}
\item the matter distribution is gravitationally bound,
\item the matter distribution has a mean stress energy $|{\mathcal T}|
\lesssim \mu^2$.
\end{itemize}
The second of these assumptions can be viewed rather flexibly if
interested only in phenomenological implications.
Regardless of whether it is satisfied, we still would expect a linearly
growing response far from the source.  To see this, recall that
the dominant piece of the solution, $f_0(y)$, from equation
(\ref{ext}) remains constant and approximately equal to one for many
orders of magnitude (for instance, $f_0(10^{-8})-f_0(10^{-3})\approx 10^{-6}$).
 Therefore, although the exterior solution would not be very reliable
near the matter source, we can be confident that at cosmic or even
intergalactic scales perturbing about de Sitter becomes appropriate
and a growing solution would still be observed.

This analysis was performed for $p=1$, but of course nothing restricts
us from considering arbitrary powers of the inverse Ricci scalar.  
To no surprise, however, varying the power only changes the
coefficient of the gravitational force leaving its qualitative
behavior alone. The instability found by Dolgov and Kawasaki
\cite{Dolgov} and the growing solution calculated in this work seem to
preclude all such theories phenomenologically. The two problems seem to
complement one another because either problem could be avoided by the 
addition of an $R^2$ term, which would not alter the cosmological solution
\cite{Nojiri2}. However, avoiding the interior instability seems to 
require the $R^2$ term to have a large coefficient, whereas avoiding the
exterior growth requires a smaller value \cite{Nojiri3}. 

None of these issues diminishes the importance that should be placed on 
considering novel approaches to understanding the dark energy problem. It 
is the responsibility of both theorists and experimentalists to construct 
and constrain candidate theories, and it is truly an exciting epoch of
human investigation for which we are just beginning to acquire these
capabilities. Greater freedom can be obtained by adding different powers 
of $R$. (Note that this generally alters the cosmological solution.) Although 
such models seem epicyclic when considered as modifications of gravity, the 
same would not be true if they were to arise from fundamental theory. For 
example, it can be shown that the brane-world scenario of Dvali, Gabadadze 
and Shifman \cite{DGS} avoids both the interior instability and the linearly 
growing force law \cite{LS}.

\begin{center}
\begin{large}
\textbf{Acknowledgements}
\end{large}
\end{center}

We thank and acknowledge James Fry for helpful correspondences during
the course of this project.  This work was partially supported by DOE 
contract DE-FG02-97ER41029 and by the Institute for Fundamental Theory.

\end{document}